# Effect of iron oxide loading on magnetoferritin structure in solution as revealed by SAXS and SANS

L. Melníková[1], V.I. Petrenko[2,3], M.V. Avdeev[2], V.M. Garamus[4], L. Almásy[5], O.I. Ivankov[2,3], L.A. Bulavin[3], Z. Mitróová[1], P. Kopčanský[1]

[1]Institute of Experimental Physics, SAS, Watsonova 47, 040 01 Kosice, Slovakia
[2]Joint Institute for Nuclear Research, Joliot-Curie 6, 141980 Dubna, Moscow region, Russia
[3]Kyiv Taras Shevchenko National University, Volodymyrska Street 64, Kyiv, 01033 Ukraine
[4]Helmholtz-Zentrum Geesthacht: Centre for Materials and Coastal Research, Max-Planck-Street 1, 21502 Geesthacht, Germany
[5]Wigner Research Centre for Physics, HAS, H-1525 Budapest, POB 49, Hungary

Corresponding author: melnikova@saske.sk, tel.: +421 55 792 2233, Fax: +421 55 633 62 92

**KEYWORDS:** magnetoferritin, apoferritin, biorelevant magnetic nanoparticles, small-angle neutron scattering, small-angle X-ray scattering, contrast variation

**ABSTRACT**

Synthetic biological macromolecule of magnetoferritin containing an iron oxide core inside a protein shell (apoferritin) is prepared with different content of iron. Its structure in aqueous solution is analyzed by small-angle synchrotron X-ray (SAXS) and neutron (SANS) scattering. The loading factor (LF) defined as the average number of iron atoms per protein is varied up to LF=800. With an increase of the LF, the scattering curves exhibit a relative increase in the total scattered intensity, a partial smearing and a shift of the match point in the SANS contrast variation data. The analysis shows an increase in the polydispersity of the proteins and a corresponding effective increase in the relative content of magnetic material against the protein moiety of the shell with the LF growth. At LFs above ~150, the apoferritin shell undergoes structural changes, which is strongly indicative of the fact that the shell stability is affected by iron oxide presence.

# 1. INTRODUCTION

Apoferritin being a part of the natural biological macromolecule of ferritin [1] represents a very useful confinement of magnetic nanoparticles inside for biomedical applications [2,3]. This almost spherical protein shell with an external diameter of 12 nm and thickness of about 2.5 nm makes it possible to disperse nanoparticles (by placing them in its cavity) in biological media and additionally minimize their possible toxic effect. It also prevents the bulk aggregation of nanoparticles and restricts their maximal size. In case of magnetic nanoparticles ($Fe_3O_4$, $\gamma$-$Fe_2O_3$) placed inside an apoferritin shell the corresponding protein is known as magnetoferritin [4]. It is of current interest for various biomedical applications, which make use of the magnetic properties of nanoparticles, such as targeted drug transport, magnetic resonance imaging, etc. [5,6]. In addition to biocompatibility another advantage of magnetoferritin is the relatively short time of synthesis, in which the magnetite ($Fe_3O_4$) is formed inside the protein cavity. Through the regulation of the iron-to-apoferritin ratio it is possible to prepare homogeneously dispersed magnetoferritin molecules with different iron oxide loading. Number of iron atoms per one molecule of protein shell is referred to as a loading factor (LF) [7].

In the previous studies the structure characterization of the magnetic core at various LFs was performed mostly by transmission electron microscopy (TEM). In particular, an increase in the magnetic nanoparticle size with the LF growth and non-spherical core shapes for low LFs were reported [7,8]. It was also found that for quite high LFs (>1000) particle aggregates are formed [7,9]. By separating and extracting the non-aggregated particles having magnetic core, the uniform magnetoferritine molecules can be crystallized to a 3D ordered magnetic array [10]. A number of experiments have been done to characterize the mineral composition of a synthetic magnetoferritin core. Thus, Mössbauer spectroscopy showed [11] that it is rather different from that of native ferritin. Faraday rotation measurements showed [12] that the composition of the core changes with increasing LF starting from maghemite with a relatively small fraction (about 10%) of magnetite at LF <1250 and varying towards 100 % of magnetite at LFs > 3250.

One can see that the previous studies of magnetoferrtin were mainly focused on the samples with LFs approaching the upper limit of the possible iron oxide content within the protein shell. However, recent investigations showed [13,14] that some structural changes of the protein shell, and also the organization of magnetoferritin in solutions are observed already at significantly lower LFs. The present paper aims at studying the influence of the magnetic content of magnetoferritin on the structure of the protein shell at low and moderate iron oxide loadings (LF < 800) by small-angle X-rays (SAXS) and neutron (SANS) scattering in order to provide additional and detailed characterization of this novel material and to follow the possible modifications of the protein cage in a wide interval of the iron oxide loading. Both kinds of the small-angle scattering technique cover the length scale of 1-100 nm, but have different sensitivity to the same elements. Especially, it concerns hydrogen whose replacement with deuterium provides wide possibilities of the so-called contrast variation in SANS. SAXS, in turn, is highly sensitive to the heavier atoms, such as iron. Here, the general size characteristics of magnetoferritin and their aggregates in aqueous solutions are first obtained by analyzing SAXS and SANS data complemented by the dynamic light scattering (DLS) measurements. Then, the SANS contrast variation is additionally applied based on the mixtures of heavy and light water to conclude about the composition of magnetoferritin.

## 2. MATERIALS AND METHODS

Natural apoferritin (*horse spleen*) was purchased from Sigma-Aldrich. Magnetoferritin with various LFs up to 800 was synthesized in anaerobic conditions at 65°C and alkaline pH as described in details elsewhere [15,16]. First, apoferritin was added into the AMPSO (3-[(1,1-Dimethyl-2-hydroxyethyl)amino]-2-hydroxy-propanesulfonic acid) buffer (0.05 M AMPSO buffered with 2 M NaOH to pH 8.6) to achieve protein concentration 6 mg·mL$^{-1}$. The buffer was deaerated for 55 minutes with nitrogen and for further 5 minutes after the addition of apoferritin. Then the solution in the reaction bottle was hermetically closed and placed in preheated water bath with temperature 65°C on a magnetic stirrer. Next, deaerated solutions of reactants, trimethylamine N-oxide and ferrous ammonium sulfate hexahydrate, were added dropwise into the reaction bottle. After the synthesis all samples were filtered through 200 nm filter to remove possible aggregates. The average loading factor of each sample was determined using UV-VIS spectrophotometer SPECORD 40 (Analytik Jena, Germany). Protein concentration was determined using the standard Bradford method at wavelength (λ) 595 nm and the amount of iron was measured after HCl/$H_2O_2$ oxidation and KSCN addition by light absorption of thiocyanate complex at λ = 400 nm.

Dynamic light scattering measurements were made on a Zetasizer Nano ZS 3600 (Malvern Instruments) at 25°C. The samples were diluted with 0.15 M NaCl to achieve the protein concentration of ~ 0.2 mg·mL$^{-1}$ and filtered through a 0.2 μm syringe filter before measurement.

For SANS contrast variation experiments magnetoferritin samples were freeze-dried for 24 hours after the synthesis to obtain a powder. 10 mg·mL$^{-1}$ solutions regarding the protein concentration were prepared by dissolving powders in $H_2O/D_2O$ mixtures with varying the $D_2O$ volume fraction. The mixtures of AMPSO buffer (0.05 M AMPSO buffered with 2 M NaOH to get pH 8.6) with the same ratios of $H_2O/D_2O$ as in the samples, were used as background solutions.

SAXS experiments were performed at the P12 BioSAXS beamline of the European Molecular Biology Laboratory (EMBL) at the storage ring PETRA III of the Deutsche Elektronen Synchrotron (DESY, Hamburg, Germany) at 20°C using a Pilatus 2M detector (1475×1679 pixels) (Dectris, Switzerland) and synchrotron radiation with a wavelength $\lambda = 0.1$ nm. The sample-detector distance was 3 m, allowing for measurements in a $q$-range of 0.11-4.4 $nm^{-1}$. The $q$-range was calibrated using the diffraction patterns of silver behenate. The experimental data were normalized to the transmitted beam intensity, corrected for a non-homogeneous detector response, and the background scattering of the aqueous buffer was subtracted. An automatic sample changer for a sample volume of 15 μL was used. The experimental time including sample loading, exposure, cleaning and drying was about 1 min per sample. The solvent scattering was measured before and after the sample scattering in order to control the eventual sample holder contamination. Four consecutive frames comprising the measurements for the solvent, the sample, and the solvent were taken. No measurable radiation damage was detected by comparing four successive time frames with 5 s exposures. The final scattering curve was obtained using the automated acquisition and analysis by averaging the scattering data collected from different frames [17].

SANS measurements were carried out at the small-angle diffractometers SANS-II at the SINQ spallation neutron source (PSI, Villigen, Switzerland) [18], operating in continuous regime, and YuMO at the IBR-2 pulsed reactor (JINR, Dubna, Russia), in time-of-flight regime. On SANS-II the scattering data were recorded at sample-detector distances of 1.3 and 4 m, with a neutron wavelength of 0.53 nm and wavelength spread of about 10%. The raw data were corrected for background, transmission and detector efficiency, and put on the absolute scale using the scattering from a 1-mm thick $H_2O$ sample, pre-calibrated by scattering from a dilute solution of polystyrene. The data were reduced by the BerSANS software package [19]. On the YuMO small-angle spectrometer a two-detector set-up with ring wire detectors were used [20]. The neutron wavelength range was 0.05-0.8 $nm^{-1}$. The measured scattering curves were corrected for background scattering from buffer solutions. For absolute calibration of the scattered intensity during the measurements a vanadium standard was used. The raw data treatment was performed by the SAS program with a smoothing mode [21]. For the measurements on both instruments the solutions were put in 1 mm thick quartz plain cells (Helma) and kept at room temperature.

## 3. RESULTS AND DISCUSSIONS

*3.1. SAXS and SANS data analysis at full contrast*

As the first step, the SAXS scattering curves for apoferritin and magnetoferritin with the minimal LF of 160 are compared in Fig. 1. The scattering curve of apoferritin (LF = 0) is well described by the form-factor of a monodisperse spherical shell [22, 23]:

$$P(q) = (1/V)^2 [V_1\Phi(qR_1) - V_2\Phi(qR_2)]^2, \qquad (1)$$

where $\Phi(x) = 3(\sin(x) - x\cos(x))/x^3$; $V_i = (4/3)\pi R_i^3$ is the volume of a sphere with radius $R_i$; $R_1 = 6.32\,(1)$ nm and $R_2 = 3.53\,(1)$ nm are the outer and inner radii of the shell, respectively, and $V = V_1 - V_2$ is the volume of the shell. The logical extension of this model for the case of magnetoferritin is the representation of the macromolecule as a spherical particle with a homogeneous, iron oxide containing core and the protein shell. However, this approach cannot describe the experimental data obtained, most likely because of modification of the magnetoferritin structure under iron oxide loading, which breaks the spherical symmetry. Therefore, the data treatment is mainly reduced to the comparative analysis of specific characteristics of the curves including the scattering invariants. Thus, in small-angle scattering at sufficiently small $q$-values one can use the Guinier approximation:

$$I(q) = I(0)\exp(-R_g^2 q^2/3) \qquad (2)$$

where the forward scattered intensity $I(0) = nV^2(\Delta\rho)^2$ is determined by the particle number density, $n$, particle volume, $V$, and the contrast, $\Delta\rho = \bar{\rho} - \rho_s$ which is the difference between the mean scattering length densities (SLDs) of the particle, $\bar{\rho}$, and solvent, $\rho_s$; and $R_g$ is the radius of gyration, the average of square distances from the center-of-mass of the macromolecule weighted by the SLD distribution. Since apoferritin, like most of proteins, is a homogeneous object in terms of the inner SLD fluctuations, its radius of gyration is strictly determined by the inner and outer radii of the protein shell, $R_g^2 = (3/5)(R_1^5 - R_2^5)/(R_1^3 - R_2^3)$, which gives $R_g = 5.25$ nm well testified by the direct approximation of Eq.2 to the experimental curve in the Gunier region ($q < 0.3$ nm$^{-1}$).

The scattering curve of magnetoferritin solution retains its character typical for a spherical shell, but an appreciable smearing of the peaks and a shift of the minima (indicated by arrows in Fig. 1) towards larger $q$-values are observed. The radius of gyration of magnetoferritin found from the Guinier approximation to the experimental curve, 4.99 nm, is slightly smaller than that of apoferritin. Also, the total intensity is larger than that for apoferritin with the same concentration of protein moiety in the solution; in particular, the forward scattered intensity of magnetoferritin exceeds that of apoferritin by 1.7 times. The observed differences cannot be attributed to a simple transformation of the hollow apoferritin shell into a core-shell structure after the cage is filled with iron oxide. First, the scattering curve of magnetoferritin cannot be properly described in terms of a simple model of monodisperse core-shell spheres as such model cannot principally explain the observed smearing. Second, the measured increase in the forward scattered intensity of magnetoferritin is too high; the volume fraction of magnetic material in the system at LF=160 is at the level of 0.005, which should give maximum a 10 % increase in the squared contrast relative to apoferritin, which is much below the observation. Since the protein shell is monodisperse, the discussed increase in the intensity suggests that the magnetic material has a non-uniform distribution over the protein shells. It was reported previously [7,8] that the loading of magnetoferritin similar to native ferritin [24,25] is characterized by some distribution of the iron content over the cages. From the viewpoint of the scattering theory one deals in this case with a distribution of $\bar\rho$ with the mean value, $\bar\rho_e$ (effective mean SLD), and width, $\sigma_p$, which determines the so-called structural polydispersity [25-27] and gives an additional contribution to the scattering. In particular, for the forward scattered intensity one can write:

$$I(0) = n\Delta\tilde\rho^2 V^2 + n\sigma_p^2 V^2, \qquad (3)$$

where the modified (for polydisperse systems) contrast is determined as $\Delta\tilde\rho = \bar\rho_e - \rho_s$. Using the experimentally found ratio between the forward scattered intensities for apoferritin and magnetoferritin at LF=160 one obtains $\sigma_p = 0.07$ e·Å$^{-3}$ (here for SLD in SAXS we use traditionally the units of number of electrons per volume). This is more than ten times larger than the difference in the mean SLD of magnetoferritin with LF=160 ($\bar\rho = 0.425$ e·Å$^{-3}$) and apoferritin ($\bar\rho = 0.42$ e·Å$^{-3}$). Thus, the volume fraction of the magnetic material in the cage varies in a much wider interval than what can be achieved for magnetoferritin regarded as a monodisperse protein cavity with just varying amount of iron oxide. This contradiction suggests that the protein shell is partly disassembled. The shell in this case is no longer a monodisperse object, and now, in addition to the structural polydispersity, the size polydispersity contributes to the scattering as well. The partial disassembling of the shell is indirectly confirmed by modelling the scattering curves by indirect Fourier transform (IFT) [28], using the GNOM program [29], which represents the scattering data in terms of the pair distance distribution function (PDD) (see inset to Fig. 1). The PDD function for magnetoferritin differs significantly from that of apoferritin, which is strongly skewed towards large distances owing to the protein shell around the empty cavity. Still, the maximal sizes are close for the two macromolecules. From the comparison of the PDD functions of magnetoferritin and a filled sphere with the diameter of apoferritin (calculated and plotted additionally in inset to Fig. 1) one can conclude that the scattering object in our case has an intermediate shape between spherical shell and sphere. This conclusion is also supported by the *ab-initio* analysis of the scattering data using the DAMMIF program [30], which models the shape of the scattering object in the homogeneous approximation by representing it with a set of sufficiently small uniform beads (Fig. 2). As compared to the scattering from apoferritin, for which DAMMIF, as expected, gives a shape very close to a hollow sphere (Fig. 2a), the DAMMIF treatment of the scattering from magnetoferritin results in a structure, which deviates strongly from a complete shell (Fig. 2b). It must be noted that this structure is some kind of an average shape, which does not exclude the existence of complete shells in the solution. The given treatment fully neglects the scattering contribution from magnetite. Still, it demonstrates clearly that the explanation of the observed shifts in the scattering minima and smearing of the curves requires quite significant deviations from a hollow sphere.

The increase in LF is accompanied by further smearing of the SAXS curves, as one can see in Fig. 3a which covers intermediate loading factors up to LF = 430. This is reflected in the PDD functions obtained by the IFT procedure (Fig. 3b) as a shift of the particle peak to smaller distances, which corresponds to a spherical symmetry violation and a transition to a more compact object. Along with it, the character of the curves changes as well, showing some specific increase in the forward scattered intensity and the radius of gyration both obtained as a result of the IFT procedure (Fig. 4). The latter is an indication of the formation of aggregates of magnetoferritin in the solutions with the LF growth. This is reflected in the corresponding PDD functions (Fig. 3b) as the appearance of a wide band above $r = 12$ nm (the expected diameter of the complete protein shell) starting from LF = 260. The ratios between the calculated and measured values of $I(0)$ and $R_g$ correspond to rather small (< 10) aggregation numbers. Such aggregation alone cannot explain the observed smearing of the curves, hence, it points to the increasing polydispersity with increasing amount of magnetic material in magnetoferritin. The size and structure polydispersity together with the absence of strictly defined scattering form-factor of the macromolecules prevent the easy determination and separation of the structure-factor which would correspond to the average effective interactions of the basic structural units (here, magnetoferritin macromolecules) in the solutions like in the case of homogeneous or multilayered structures [31, 32]. Instead, the structures formed resemble more the partly aggregated particles in aqueous dispersions of magnetite nanoparticles coated with surfactant shells [33].

The SANS curves (Fig. 5a) and the corresponding PDD functions (Fig. 5b) for another series of magnetoferritin solutions cover a more extended interval over LFs, up to LF = 800. For the intermediate LFs (LF < 600) the similar treatment generally repeats the previous conclusions of the SAXS analysis; yet, the aggregate effect starts to be visible at higher LF and is characterized by smaller aggregate size for the second series. At the same time, starting from LF = 600 a tendency towards a sharp increase in the aggregation is seen, which is well distinguished as a drastic widening of the corresponding PDD functions (Fig. 5b). A further increase in LF would make it impossible to treat the curves in the same way at the given instrumental resolution, which is determined by the minimal $q$-value corresponding to the detectable maximal size of the scattering objects. The formation of stable aggregates in magnetoferritin solutions with the LF growth is confirmed by the DLS measurements from diluted solutions. In Fig. 6 the LF-dependences of the mean hydrodynamic radius, $<R_{hydr}>$, and of radius of gyration obtained by SAXS and SANS are compared. For apoferritin the $<R_{hydr}>$-value is fully consistent with the small-angle scattering data if one takes into account that in this case, the radius of gyration corresponds to the radius of the hollow protein shell of about 6 nm, and the hydrodynamic radius naturally exceeds this value by about 10%. A non-monotonic size growth is revealed in the three kinds of experiments, as shown in Fig. 6 from which one can reliably conclude that a tendency to a slight aggregation of magnetoferritin is seen at LF over the interval of 160 – 510, and the aggregation becomes more intensive at LF above 600. The discussed LF-dependences do not fully repeat themselves most probably because of a strong sensitivity to stochastic factors during the sample preparation (e.g. intensity and time of solution deaeration and stirring), which is typical for liquid dispersions of nanoparticles.

It is interesting to compare the size characteristics of magnetoferritin with those of ferritin (Figs. 3, 5, 6) in solutions under the same conditions. The natural LF-values of ferritin (the core has a ferrihydrite-like structure) are close to LF = 2000. One can see that despite the large iron content the scattering curves from ferritin show more pronounced oscillating behaviour, thus reflecting rather high monodispersity and structural stability of this macromolecule. At the same time, the corresponding PDD functions (Figs. 3 b, 5 b) indicate that the ferritin solutions are not free of some small aggregates; still, their mean size is significantly lower as compared to the solutions of magnetoferritin.

The disassembling of the protein shell in apoferritin can take place under some conditions, in particular in strongly acidic solutions [34]. Thus, it was shown that the complete disappearance of the characteristic peaks in SAXS curves from disassembled apoferritin strictly takes place when 12 out of the 24 structural units are removed from the shell. In our experiments pH was kept constant at 8.6, which is optimal for the stability of apoferritin structure, but the character of the observed smearing of the scattering curves was the same, thus indicating that in average about half of the apoferritin shell in magnetoferritin is destroyed when LF approaches 1000.

*3.2. SANS contrast variation*

The contrast variation technique in SANS experiments on moderately polydisperse objects makes it possible to conclude about the polydispersity degree in terms of the weighted averaged scattering length density distribution over the studied particles [26]. For this purpose the scattering from the system under study is analysed, varying the content of a deuterated component of the solvent. Here, the SANS contrast variation data based on substitution of light ($H_2O$) for heavy ($D_2O$) water is used to conclude about the change in the polydispersity with rising LF. The samples with low (LF = 160) and relatively high (LF = 510) loading factors were investigated. The upper LF-value was chosen to avoid the large aggregation which starts, as shown above, at LF of about 600. The Guinier region for the different contrasts (Fig. 7) was used to determine the $I(0)$ parameter according to Eq. 1. $I(0)$ as a function of the volume fraction of $D_2O$ is shown in Fig. 8; its minimum for the case of polydisperse particles corresponds to the effective match point [26]. At the considered LF-values the additional magnetic neutron scattering contribution can be neglected. The upper estimates give for LF=510 that its contribution to the forward scattered intensity is less than 2%. Already in the monodisperse approximation, under the assumption that the magnetic core in magnetoferritin consists of magnetite ($Fe_3O_4$, SLD = $6.9 \cdot 10^{10}$ cm$^{-2}$), the shifts of the effective match points (the corresponding SLDs are $2.46 \cdot 10^{10}$ cm$^{-2}$ and $2.79 \cdot 10^{10}$ cm$^{-2}$ for LF=160 and LF=510, respectively) as compared to the protein moiety of apoferritin (SLD $2.34 \cdot 10^{10}$ cm$^{-2}$) give 0.026 and 0.099, respectively, for the volume fractions of magnetic material in the protein cage. These values are much larger than the amount of iron loaded during the synthesis (0.005 and 0.017 for LF=160 and LF=510, respectively). Assuming maghemite ($Fe_2O_3$, SLD = $6.7 \cdot 10^{10}$ cm$^{-2}$) in the magnetic core, the result would differ by less than 5%.

The obtained match points are significantly higher than those expected for a core filled with iron oxides. Therefore, the SANS contrast variation strongly points to an abnormally high average ratio between the content of the magnetic material and protein, which can be explained by the partial disassembling of the shell, leading to an effective growth of the relative content of the magnetic component in the structure of magnetoferritin. The residual scattering in the effective match points, which is an indicator of polydispersity, increases, thus explaining the broadening of the polydispersity for larger LFs. This is consistent with the smearing of the scattering curves with the LF growth in Figs. 3a, 5a.

It should be noted that despite the concluded disassembling of the protein shell the magnetoferritins remain soluble as a whole. Also, the solutions themselves are stable for at least three months, without signs of precipitates. The mechanism of the effect of magnetic loading on the protein structure is unclear. As mentioned previously, apoferritin disassembly was observed at pH below 3.4 [34], which, however, is not our case, since magnetoferritin was prepared in alkaline pH 8.6 and anaerobic conditions. While it is not possible to control pH directly during the synthesis process, after the synthesis the pH value was checked and only slight decrease for higher LFs was detected, remaining above pH 7 for all LFs.

One can relate the observed structural change of the protein with a specific effect of magnetic nanoparticles placed in the cavity on the protein shell. So far there is no general understanding of interactions between nanoparticles and proteins despite of the extensive studies of this problem in recent years. In particular, the interaction of various nanoparticles with specific protein aggregates (amyloids) can be mentioned. Among different types of probed materials [35] magnetic nanoparticles of iron oxides show inhibiting and even disaggregating effect on amyloidal aggregation [36-40].

**CONCLUSIONS**

In summary, the combined SAXS/SANS analysis complemented by DLS measurements of magnetoferritin aqueous solutions at loading factors in the interval of 160 – 800 reveals two competitive effects when increasing the LF. First, a partial disassembling of apoferritin shell in magnetoferritin, starting from the smallest of the applied LFs is observed. The effect increases with the LF growth and, in addition to the structure polydispersity (distribution of loading over the proteins), results in a moderate size polydispersity of magnetoferritin. Second, at LFs above 160 a tendency towards slight aggregation (aggregation number below 10) of magnetoferritin is observed; it takes place in a wide protein concentration interval of 0.2 – 20 mg ml$^{-1}$ but is rather sensitive to the preparation procedure. The aggregation becomes more intensive at LFs above 600.

**ACKNOWLEDGEMENTS**

This work was supported by the Slovak Scientific Grant Agency VEGA (projects No. 0041, 0045), by the European Structural Funds, projects NANOKOP No. 26110230061 and 26220120021, PHYSNET No. 26110230097, PROMATECH No. 26220220186, APVV 0171–10 (METAMYLC) M-ERA.NET MACOSYS and 226507-NMI3. The kind support from Urs Gasser at the SANS II instrument (PSI) and Manfred Roessle at the P12 BioSAXS beamline (EMBL/DESY, Petra III) is acknowledged. This work is based on experiments performed at the Swiss spallation neutron source SINQ, Paul Scherrer Institut, Villigen, Switzerland. L.A. thanks the Hungarian Scholarship Board for the support of a short research stay at the IEP SAS, and acknowledges the financial support from project KMR12-1-2012-0226 granted by the National Development Agency (NDA) of Hungary.

**REFERENCES**

[1] E.C. Theil, R.K. Behera, T. Tosha, Coord. Chem. Rev. 257 (2013) 579.
[2] N. Galvez, B. Fernandez, E. Valero, P. Sanchez, R. Cuesta, J.M. Dominguez-Vera, C.R. Chimie 11 (2008) 1207.
[3] M. Ceolin, N. Galvez, P. Sanchez, B. Fernandez, J.M. Dominguez-Vera, Eur. J. Inorg. Chem. 2008 (2008) 795.
[4] F.C. Meldrum, B.R. Heywood, S. Mann, Science 257 (1992) 522.
[5] M. Uchida, M.L. Flenniken, M. Allen , D.A. Willits, B.E. Crowley, S. Brumfield, A.F. Willis, L. Jackiw, M. Jutila, M.J. Young, T. Douglas, J. Am. Chem. Soc. 128 (2006) 16626.
[6] K. Fan, Ch. Cao, Y. Pan, D. Lu, D. Yang, J. Feng, L. Song, M. Liang, X. Yan, Nat. Nanotechnol. 7 (2012) 459.
[7] K.K.W. Wong, T. Douglas, S. Gider, D.D. Awschalom, S. Mann, Chem. Mater. 10 (1998) 279.
[8] M.J. Martinez-Perez, R. de Miguel, C. Carbonera, M. Martinez-Julvez, A. Lostao, C. Piquer, C. Gomez-Moreno, J. Bartolome, F. Luis, Nanotechnology 21 (2010) 465707.
[9] Z. Mitroova, L. Melnikova, J. Kovac, M. Timko, P. Kopcansky, Acta Phys. Pol A. 121 (2012) 1318.
[10] O. Kasyutich, A. Sarua, W. Schwarzacher, J. Phys. D: Appl. Phys. 41 (2008) 134022.
[11] D.P.E. Dickson, S.A. Walton, S. Mann, K. Wong, Nanostruct Mater. 9 (1997) 595.
[12] M. Koralewski, J.W. Kłos, M. Baranowski, Z. Mitroova, P. Kopcansky, L. Melnikova, M. Okuda, W. Schwarzacher, Nanotechnology 23 (2012) 355704.
[13] L.Melnikova, Z.Mitroova, M.Timko, J.Kovač, M. Koralewski, M.Pochylski, M.V.Avdeev, V.I.Petrenko, V.M. Garamus, L.Almasy, P.Kopčanský, Magnetohydrodynamics 48 (2013) 407.
[14] L.Melníková, Z.Mitróová, M.Timko, J.Kováč, M.V.Avdeev, V.I.Petrenko, V.M. Garamus, L.Almásy, P. Kopčanský, Mendeleev Comm. 24 (2014) 80.
[15] M. Koralewski, M. Pochylski, Z. Mitroova, M. Timko, P. Kopcansky, L. Melnikova, J. Magn. Magn. Mater. 323 (2011) 2413.
[16] M.T. Klem, M. Young, T. Douglas, Mater. Today 8 (2005) 28.
[17] D. Franke, A.G. Kikhney, D.I. Svergun, Nucl. Instrum. Methods A 689 (2012) 52.
[18] P. Strunz, K. Mortensen, S. Janssen, Phys. B: Condens. Matter 350 (2004) E783.

[19]  U. Keiderling, Appl. Phys. A 74 (2002) S1455.
[20]  A.I. Kuklin, A.Kh. Islamov, V.I. Gordeliy, Neutron News 16 (2005) 16.
[21]  A.G. Soloviev, T.N. Murugova, A.H. Islamov, A.I. Kuklin, J. Phys. Conf. Ser. 351 (2012) 012027.
[22]  W. Häußler, A. Wilk, J. Gapinski, A. Patkowski, J. Chem. Phys., 117 (2002).
[23]  A.I. Kuklin, T.N. Murugova, O.I. Ivankov, A.V Rogachev, D.V. Soloviov, Y S Kovalev, A.V. Ishchenko, A. Zhigunov, T.S. Kurkin, V.I. Gordeliy, J. Physics: Conf. Ser. 351 (2012) 012009.
[24]  H.B. Stuhrmann, E.D. Duee, J. Appl. Cryst. 8 (1975) 538.
[25]  H.B. Stuhrmann, J. Haas, K. Ibel, M.H.J. Koch, R.R. Crichton, J. Mol. Biol. 100 (1976) 399.
[26]  M.V. Avdeev, J. Appl. Cryst. 40 (2007) 56.
[27]  M.V. Avdeev, V.L. Aksenov, Phys. Usp. 53 (2010) 971.
[28]  O. Glatter, J. Appl. Cryst. 10 (1977) 415.
[29]  D.I. Svergun, A.V. Semenyuk, L.A. Feigin, Acta Cryst. A44 (1988) 244.
[30]  D. Franke, D.I. Svergun, J. Appl. Cryst. 42 (2009) 342.
[31]  J. Skov Pedersen, Adv. Coll. Inter. Sci. 70 (1997) 171.
[32]  C.L.P. Oliveira, B.B. Gerbelli, E.R.T. Silva, F. Nallet, L. Navailles, E.A. Oliveira, J.S. Pedersen, J. Appl. Cryst. 45 (2012) 1278.
[33]  M.V. Avdeev, B. Mucha, K. Lamszus, L. Vekas, V.M. Garamus, A.V. Feoktystov, O. Marinica, R. Turcu, R. Willumeit, Langmuir 26 (2010) 8503.
[34]  M. Kim, Y. Rho, K.S. Jin, B. Ahn, S. Jung, H. Kim, M. Ree, Biomacromolecules 12 (2011) 1629.
[35]  W. Wu, X. Sun, Y. Yu, J. Hu, L. Zhao, Q. Liu, Y. Zhao, Y. Li, Biochem. Biophys. Res. Comm. 373 (2008) 315.
[36]  K. Siposova, M. Kubovcikova, Z. Bednarikova, M. Koneracka, V. Zavisova, A. Antosova, P. Kopcansky, Z. Daxnerova, Z. Gazova, Nanotechnology 23 (2012) 055101.
[37]  A. Bellova, E. Bystrenova, M. Koneracka, P. Kopcansky, F. Valle, N. Tomasovicova, M. Timko, J. Bagelova, F. Biscarini, Z. Gazova, Nanotechnology 21 (2010) 065103.
[38]  M. Mahmoudi, F. Quinlan-Pluck, M.P. Monopoli, S. Sheibani, H. Vali, K.A. Dawson, I. Lynch, ACS Chem. Neurosci. 4 (2013) 475.
[39]  L. Xiao, D. Zhao, W.-H. Chan, M.M.F. Choi, H.-W. Li, Biomaterials 31 (2010) 91.
[40]  C. Cabaleiro-Lago, F. Quinlan-Pluck, I. Lynch, K.A. Dawson, S. Linse, ACS Chem. Neurosci. 1 (2010) 279.

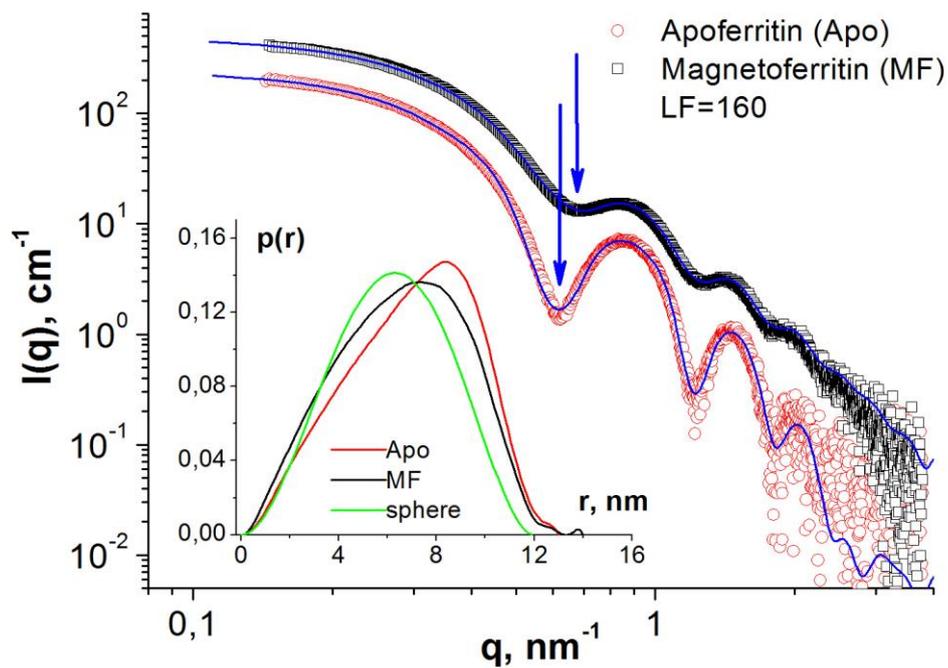

**Fig. 1.** Experimental SAXS curves for apoferritin and magnetoferritin with low LF. Smearing and shift of minima are indicated by arrows against the first minimum. Solid lines correspond to the model curves obtained by DAMMIF (see text). Relative experimental errors at $q < 1.8$ nm$^{-1}$ do not exceed 1%. Inset shows PDD functions of apoferritin and magnetoferritin (results of the IFT treatment of the experimental curves) and a sphere with radius of 6 nm (model calculations).

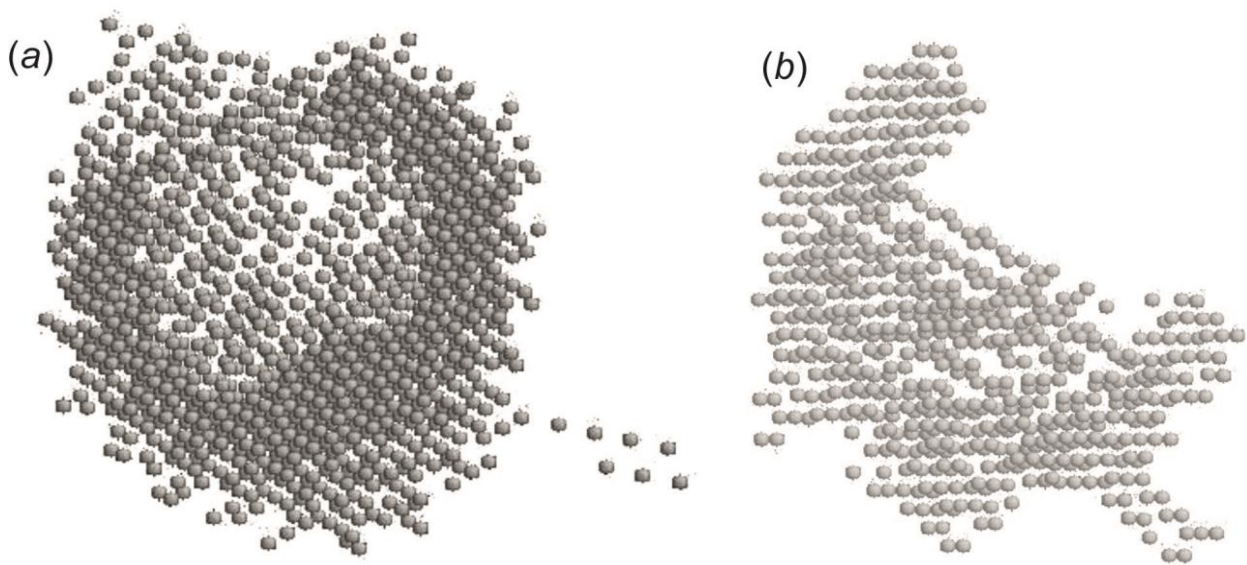

**Fig. 2.** Bead models of apoferritin (*a*) and magnetoferritin with LF=160 (*b*) obtained by the DAMMIF procedure on the data shown in Fig. 1.

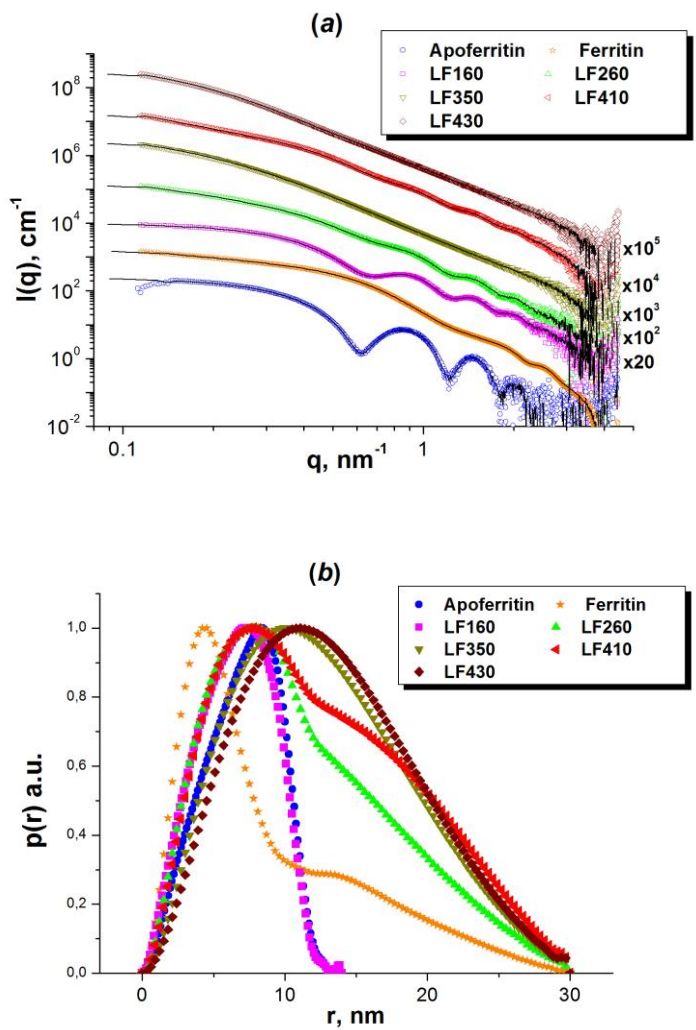

**Fig. 3.** SAXS data for magnetoferritin with different LF (intermediate values are covered) and comparison with apoferritin (LF = 0) and ferritin (LF = 1990): experimental scattering curves (*a*) and PDD functions as a result of IFT treatment (*b*). The relative experimental errors in the scattering data points at $q < 1.8$ nm$^{-1}$ do not exceed 1%. The concentrations of proteins are 2.35 mg·mL$^{-1}$ for apoferritin, 44 mg·mL$^{-1}$ for ferritin, and 2.81 mg·mL$^{-1}$ for magnetoferritin LF160, 7.18 mg·mL$^{-1}$ LF260, 6.96 mg·mL$^{-1}$ LF350, 7.95 mg·mL$^{-1}$ LF410, 8.14 mg·mL$^{-1}$ LF430. For convenient view the curves in (*a*) are shifted vertically by multiplying by the factor indicated at the right. In (*a*) the solid lines show the IFT fits.

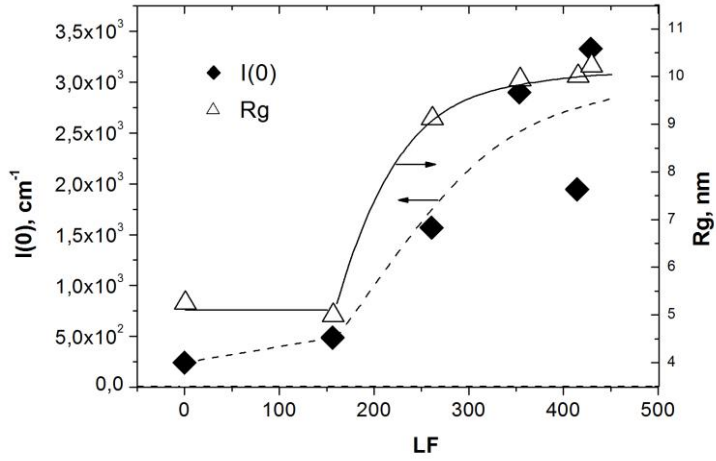

**Fig. 4.** LF-dependences of *I*(0) and $R_g$ found by the IFT treatment of the SAXS experimental curves in Fig. 3a. The lines are plotted to follow the tendencies. Experimental errors do not exceed the size of the points.

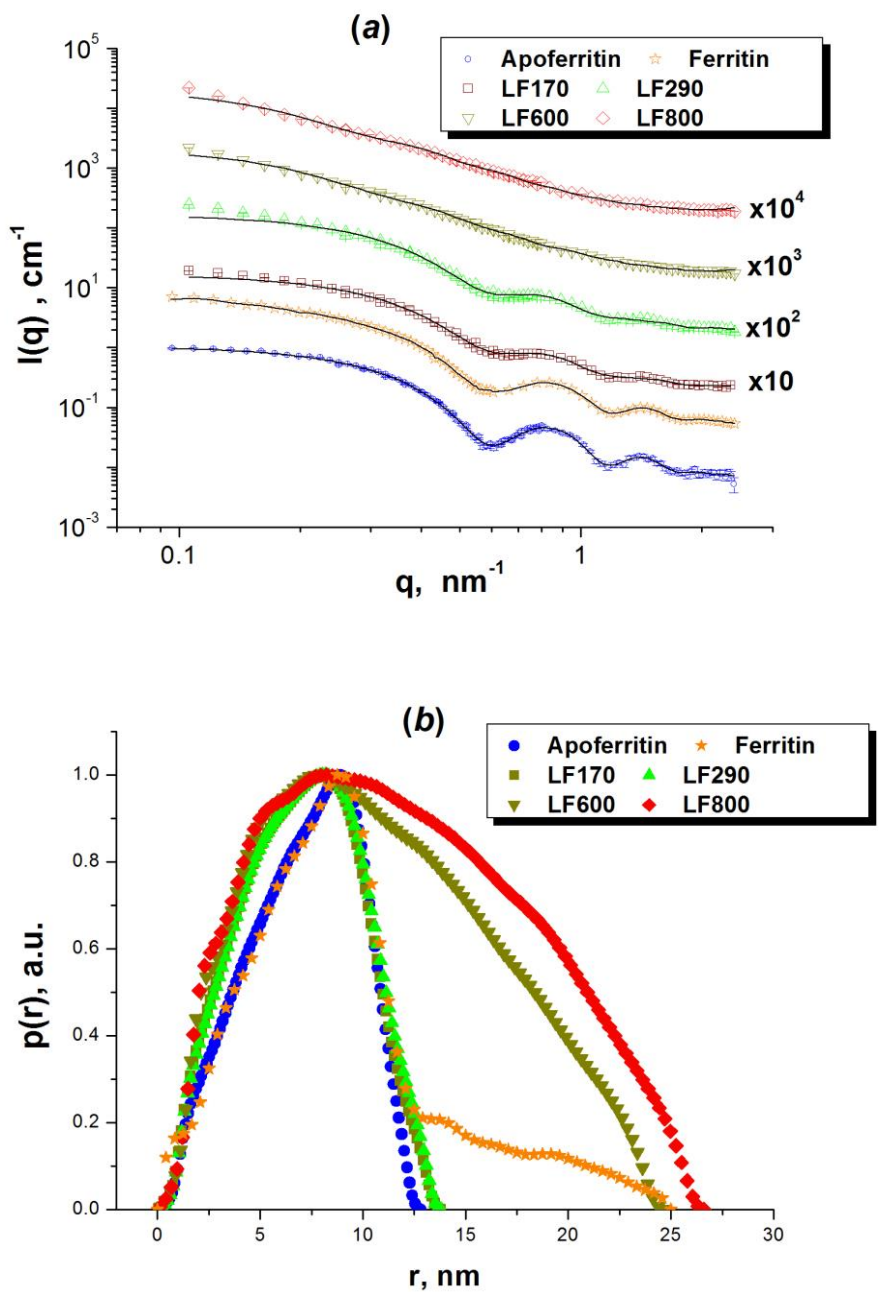

**Fig. 5.** SANS data for magnetoferritin with different LF (high values are covered) and comparison with apoferritin (LF = 0) and ferritin (LF = 1990): experimental scattering curves (*a*) and PDD functions as a result of the IFT treatment (*b*). In (*a*) the relative experimental errors do not exceed 5 %. The protein concentration in all magnetoferritin solutions based on $D_2O$ is 20 mg·mL$^{-1}$. The concentrations of proteins are 2.35 mg·mL$^{-1}$ in apoferritin and 44 mg·mL$^{-1}$ in ferritin solutions. For convenient view data in (*a*) are shifted vertically by multiplying by the indicated factor. In (*a*) the solid lines show the IFT fits.

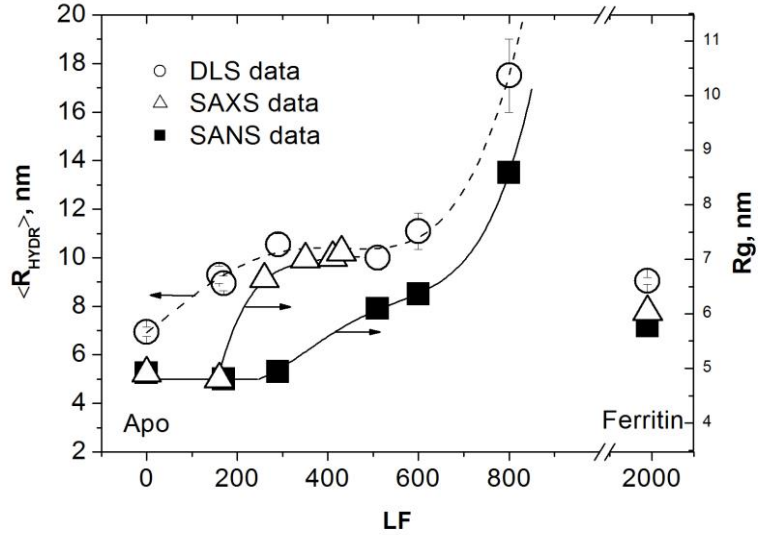

**Fig. 6.** LF-dependences of hydrodynamic radius (DLS data) and radius of gyration (SAXS and SANS data, IFT treatment) of magnetoferritin for different series of samples and comparison with the corresponding values of apoferritin (Apo) and ferritin. For SAXS and SANS data the experimental errors do not exceed the size of the points. The lines are plotted to follow the tendencies.

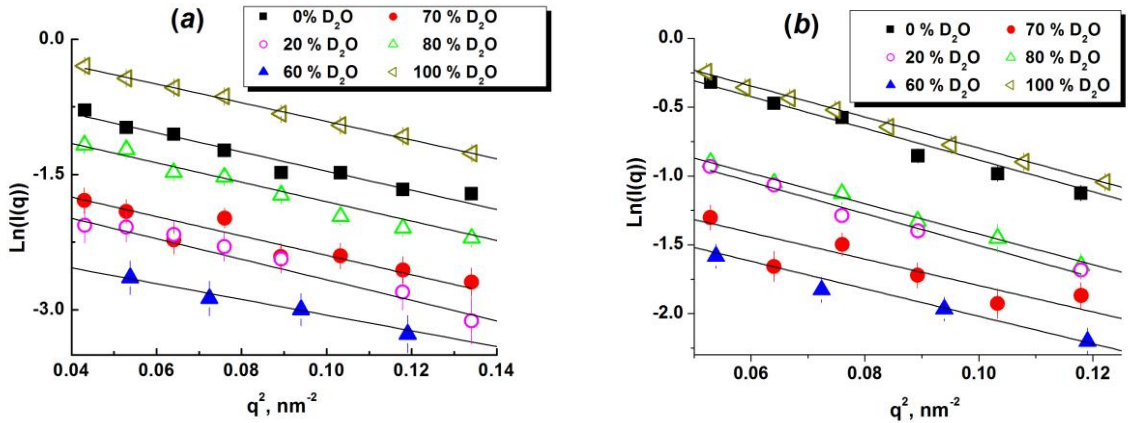

**Fig. 7.** Guinier plots for SANS contrast variation data for magnetoferritin solutions with LF 160 (*a*) and LF 510 (*b*).

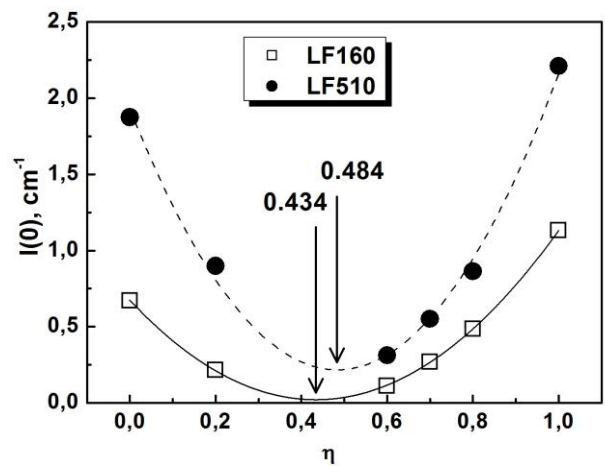

**Fig. 8.** The change in the forward scattered intensity $I(0)$ for two samples of magnetoferritin (loading factors LF=160 and LF=510) solutions with varying volume fraction of $D_2O$, $\eta$, in the solvent. The experimental errors do not exceed the size of the points. Effective match points corresponding to the intensity minima are indicated by vertical arrows.